\documentclass[aps,floatfix,preprint,preprintnumbers,nofootinbib,superscriptaddress,natbib]{revtex4}

\usepackage{graphicx,float}
\usepackage{epsfig}		
\usepackage{dcolumn}
\usepackage{bm}
\usepackage{subfigure}

\usepackage[all]{xy}
\usepackage{amsmath,upgreek}
\usepackage{amssymb}

\usepackage{pdfpages}
\usepackage{color}
\usepackage{graphicx,epstopdf}
\usepackage[colorlinks=true,
 linkcolor=red,
 urlcolor=purple,
 citecolor=blue,hyperindex]{hyperref}
\def\0{\mbox{\tiny $0$}}
\def\1{\mbox{\tiny $1$}}
\def\2{\mbox{\tiny $2$}}
\def\3{\mbox{\tiny $3$}}
\def\4{\mbox{\tiny $4$}}
\def\5{\mbox{\tiny $5$}}
\def\6{\mbox{\tiny $6$}}
\def\7{\mbox{\tiny $7$}}
\def\8{\mbox{\tiny $8$}}
\def\9{\mbox{\tiny $9$}}

\def\f14{\mbox{\tiny $\frac{1}{4}$}}

\begin{document}

\title{Phase-space gaussian ensemble quantum camouflage}

\renewcommand{\baselinestretch}{1.2}
\author{Alex E. Bernardini}
\email{alexeb@ufscar.br}
\affiliation{~Departamento de F\'{\i}sica, Universidade Federal de S\~ao Carlos, PO Box 676, 13565-905, S\~ao Carlos, SP, Brasil.}
\author{O. Bertolami}
\email{orfeu.bertolami@fc.up.pt}
\affiliation{Departamento de F\'isica e Astronomia, Faculdade de Ci\^{e}ncias da
Universidade do Porto, Rua do Campo Alegre 687, 4169-007, Porto, Portugal.}
\altaffiliation{Also at Centro de F\'isica do Porto, Rua do Campo Alegre 687, 4169-007, Porto, Portugal.} 
\date{\today}

\begin{abstract}
Extending the phase-space description of the Weyl-Wigner quantum mechanics to a subset of non-linear Hamiltonians in position and momentum, gaussian functions are identified as the quantum ground state.
Once a Hamiltonian, $H^{W}(q,\,p)$, is constrained by the $\partial ^2 H^{W} / \partial q \partial p = 0$ condition, flow properties for generic $1$-dim systems can be analytically obtained in terms of Wigner functions and Wigner currents.
For gaussian statistical ensembles, the exact phase-space profile of the quantum fluctuations over the classical trajectories are found, so to interpret them as a suitable Hilbert space state configuration for confronting quantum and classical regimes.
In particular, a sort of {\em quantum camouflage} where the stationarity of classical statistical ensembles can be camouflaged by the stationarity of gaussian quantum ensembles is identified.
Besides the broadness of the framework worked out in some previous examples, our results provide an encompassing picture of quantum effects on non-linear dynamical systems which can be interpreted as a first step for finding the complete spectrum of  non-standard Hamiltonians.
\end{abstract}


\date{\today}
\maketitle

\section{Introduction}

The Weyl-Wigner (WW) \cite{Wigner,Ballentine,Case,Meu2018} phase-space formalism of quantum mechanics (QM) encompasses the dynamics of quantum systems and offers an equivalent description of QM in terms of {\em quasi}-probability distribution functions of position and momentum coordinates. The formalism provides subtle insights about boundaries between quantum and classical physics as well as a straightforward access to quantum information issues \cite{Neumann,Zurek01,Zurek02,Steuernagel3}.
Our proposal in this work is to explore the WW formulation of the QM extended to generic Hamiltonian systems  described by a Weyl transformed Hamiltonian of the form, 
\begin{equation}
H^W(q,\,p) = K^W(p) + V^W(q),
\label{nlh}
\end{equation}
constrained by the $\partial ^2 H^W / \partial q \partial p = 0$ condition, where $K^W(p)$ and $V^W(q)$ are arbitrary operator functions of $p$ and $q$, respectively.

As an extension of previous results obtained for Harper-like systems \cite{Novo21A} and quantized prey-predator dynamics \cite{Novo21B,Novo21B2}, our investigation here is concerned with the identification of gaussian ensembles as quantum ground states of a particular class of Hamiltonians, Eq.~\eqref{nlh}.

The outline of our manuscript is as follows.
Sec. II is concerned with the fundamentals of the extended WW framework.
In Sec. III, results are specialized for gaussian ensembles.
This analysis shows that the Wigner flow framework provides the overall quantum distortion over the phase-space classical pattern, where the quantum effects are analytically computed through a convergent infinite series expansion in terms of quadratic powers of the Planck constant \cite{Novo21A}.
In Sec. IV, a subtle aspect involving the classical-quantum correspondence allows one to identify a kind of phase-space {\em quantum camouflage} involving gaussian states, i.e. the stationarity of classical statistical ensembles can be camouflaged by the stationarity of gaussian quantum ensembles.
Our conclusions are presented in Sec. V, and is geared towards a broader understanding of quantum-like effects on non-linear dynamical systems.

\section{Extended Weyl-Wigner framework}

Supported by the Heisenberg-Weyl algebra, which at $1$-dim is driven by the position-momentum commutation relation, $[q,\,p] = i\,\hbar$, the Wigner phase-space {\em quasi}-distribution function, $W(q,\, p)$, allows for a broader interpretation of the QM framework, when compared to either Schr\"odinger or Heisenberg pictures.
Defined through the Weyl transform of the density matrix operator, $\hat{\rho} = |\psi \rangle \langle \psi |$,
the Wigner function is given by
\begin{equation}
 (2\pi \hbar)^{-1} \hat{\rho} \to W(q,\, p) = (\pi\hbar)^{-1} 
\int^{+\infty}_{-\infty} \hspace{-.35cm}ds\,\exp{\left[2\, i \, p \,s/\hbar\right]}\,
\psi(q - s)\,\psi^{\ast}(q + s),\label{222}
\end{equation}
which was originally introduced in the context of accounting for quantum corrections to TD equilibrium states \cite{Wigner}, akin to the formalism of statistical mechanics.

The WW phase-space formulation covers all the QM paradigms \cite{Ballentine,Case,Meu2018}, where the Wigner function dynamical properties, $W(q,\,p) \to W(q,\,p;\,t)$, are connected to the Hamiltonian dynamics by means of a vector flux \cite{Steuernagel3,NossoPaper,Meu2018}, $\mathbf{J}(q,\,p;\,t)$, decomposed into the phase-space coordinate directions, $\hat{q}$ and $\hat{p}$, as $\mathbf{J} = J_q\,\hat{q} + J_p\,\hat{p}$, so to 
reproduce a flow field connected to the Wigner function dynamics through the continuity equation \cite{Case,Ballentine,Steuernagel3,NossoPaper,Meu2018},
\begin{equation}
{\partial_t W} + {\partial_q J_q}+{\partial_p J_p} =0,
\label{alexquaz51}
\end{equation}
where the shortened notation for partial derivatives is set as $\partial_a \equiv \partial/\partial a$. In this case, for a non-relativistic Hamiltonian operator, ${H}({Q},\,{P})$, from which the Weyl transform yields 
\begin{equation}
{H}({Q},\,{P}) = \frac{{P}^2}{2m} + V({Q}) \quad\to \quad H^{W}(q,\, p) = \frac{{p}^2}{2m} + V(q),
\end{equation}
one has \cite{Case,Ballentine,Steuernagel3,NossoPaper}
\begin{equation}
J_q(q,\,p;\,t)= \frac{p}{m}\,W(q,\,p;\,t), \label{alexquaz500BB}
\end{equation}
and
\begin{equation}
J_p(q,\,p;\,t) = -\sum_{\eta=0}^{\infty} \left(\frac{i\,\hbar}{2}\right)^{2\eta}\frac{1}{(2\eta+1)!} \, \left[\partial_q^{2\eta+1}V(q)\right]\,\partial_p ^{2\eta}W(q,\,p;\,t),
\label{alexquaz500}
\end{equation}
with $\partial^s_a \equiv (\partial/\partial a)^s$, from which one notices that the above identified series expansion contributions from $\eta \geq 1$ introduce quantum modifications which are reflected onto the phase-space trajectories. In fact, from Eq.~\eqref{alexquaz500}, one sees that the suppression of the $\eta \geq 1$ contributions results into a classical Hamiltonian description of the phase-space probability distribution dynamics in terms of classical (Liovillian equivalent) Wigner currents,
\begin{equation}
J^{\mathcal{C}}_q(q,\,p;\,t)= +({\partial_p H^{W}})\,W(q,\,p;\,t), \label{alexquaz500BB2}
\end{equation}
and
\begin{equation}
J^{\mathcal{C}}_p(q,\,p;\,t) = -({\partial_q H^{W}})\,W(q,\,p;\,t),
\label{alexquaz500CC2}
\end{equation}
which, once substituted into the Eq.~\eqref{alexquaz51}, deliver back the Liouville equation. The classical phase-space velocity is identified by $\mathbf{v}_{\xi(\mathcal{C})} = \dot{\mbox{\boldmath $\xi$}} = (\dot{q},\,\dot{p})\equiv ({\partial_p H^{W}},\,-{\partial_q H^{W}})$, with $\mbox{\boldmath $\nabla$}_{\xi}\cdot \mathbf{v}_{\xi(\mathcal{C})}= \partial_q \dot{q} + \partial_p\dot{p} = 0$, where over {\em dots} denote the time derivative, $d/dt$.
Likewise, for a quantum current parameterized by $\mathbf{J} = \mathbf{w}\,W$, where the Wigner phase-space velocity, $\mathbf{w}$, is the quantum analog of $\mathbf{v}_{\xi(\mathcal{C})}$, a suitable divergent behavior is identified by
\begin{equation}
\mbox{\boldmath $\nabla$}_{\xi} \cdot \mathbf{w} = \frac{W\, \mbox{\boldmath $\nabla$}_{\xi}\cdot \mathbf{J} - \mathbf{J}\cdot\mbox{\boldmath $\nabla$}_{\xi}W}{W^2},
\label{zeqnz59}
\end{equation}
since $\mbox{\boldmath $\nabla$}_{\xi}\cdot\mathbf{J} = W\,\mbox{\boldmath $\nabla$}_{\xi}\cdot\mathbf{w}+ \mathbf{w}\cdot \mbox{\boldmath $\nabla$}_{\xi}W$ \cite{Steuernagel3}.

Turning to Hamiltonians in the form of Eq.~\eqref{nlh}, our departing point \cite{Novo21A} has been the Von Neumann equation for the state density operator, $\hat{\rho} = \vert \psi \rangle \langle \psi\vert$, obtained in Ref.~ \cite{Ballentine}
\begin{equation}
\partial_t\hat{\rho} = i\hbar^{-1} \left[\hat{\rho}, \, H \right] \equiv{\partial_t^{^{(K)}}\hat{\rho}} ~ + ~{\partial_t^{^{(V)}}\hat{\rho}},\quad \mbox{with} ~~ {\partial_t^{^{( \mathcal{A})}}\hat{\rho}} = i\hbar^{-1} \left[\hat{\rho}, \, \mathcal{A}\right],
\label{dens}
\end{equation}
which can then be separately evaluated in momentum and position representations, for $\mathcal{A} \equiv K({P}), \,V({Q})$.
Hence using the Wigner function properties from Eq.~\eqref{222} to transform each contribution into its respective Wigner representation (cf. Ref.~\cite{Ballentine} for non-relativistic QM, one has:
\begin{eqnarray}
\partial^{^{(K)}}_t\langle p \vert {\rho} \vert p'\rangle &=& i \hbar^{-1}\langle p \vert {\rho} \vert p'\rangle
\,\left(K(p') - K(p)\right)\Rightarrow\\
 \partial^{^{(K)}}_t W(q,\,p;\,t) &=& i \hbar^{-1} (\pi\hbar)^{-1}
\int^{+\infty}_{-\infty} \hspace{-.35cm}dr\,
\rho^{W,\varphi}_{(p-r;\,p+r)}
\exp{\left[-2\, i \, q \,r/\hbar\right]}
\,\left[K(p+r) - K(p-r)\right],\nonumber
\label{Wigner222BB}
\end{eqnarray}
where $\rho^{W,\varphi}_{(p-r;\,p+r)} \equiv \langle p-r \vert {\rho} \vert p+r\rangle$ corresponds to $\varphi(p- r)\,\varphi^{\ast}(p+ r)$, and
\begin{eqnarray}
\partial^{^{(V)}}_t\langle q \vert {\rho} \vert q'\rangle &=& i \hbar^{-1}\langle q \vert {\rho} \vert q'\rangle
\,\left[V(q') - V(q)\right]\Rightarrow\\
 \partial^{^{(V)}}_t W(q,\,p;\,t) &=& i \hbar^{-1} (\pi\hbar)^{-1}
\int^{+\infty}_{-\infty} \hspace{-.35cm}ds\,
\rho^{W,\psi}_{(q-s;\,q+s)}
\exp{\left[2\, i \, p \,s/\hbar\right]}
\,\left(V(q+s) - V(q-s)\right),\nonumber
\label{Wigner222CC}
\end{eqnarray} 
where $\rho^{W,\psi}_{(q-s;\,q+s)} \equiv \langle q - s \vert {\rho} \vert q + s\rangle$ corresponds to $\psi(q - s)\,\psi^{\ast}(q + s)$\footnote{From, $W(q,\, p)$, marginal distributions which return position and momentum distributions upon integrations over the momentum and position coordinates are, respectively,
\begin{equation}
\vert \psi(q)\vert^2 = \int^{+\infty}_{-\infty} \hspace{-.35cm}dp\,W(q,\, p)
\qquad
\leftrightarrow
\qquad
\vert \varphi(p)\vert^2 = \int^{+\infty}_{-\infty} \hspace{-.35cm}dq\,W(q,\, p),\nonumber
\end{equation}
such that the associated Fourier transform sets
\begin{equation}
 \varphi(p)=
(2\pi\hbar)^{-1/2}\int^{+\infty}_{-\infty} \hspace{-.35cm} dq\,\exp{\left[i \, p \,q/\hbar\right]}\,
\psi(q).\nonumber
\end{equation}}.
Now, by noticing that
\begin{equation}
K(p+r) - K(p-r) = 2\sum_{\eta=0}^{\infty}\frac{r^{2\eta+1}}{(2\eta+1)!} \,\partial_p^{2\eta+1}K(p),
\label{alexquaz500BB}
\end{equation}
and
\begin{equation}
V(q+s) - V(q-s) = 2\sum_{\eta=0}^{\infty}\frac{s^{2\eta+1}}{(2\eta+1)!} \,\partial_q^{2\eta+1}
V(q),
\label{alexquaz500CC}
\end{equation}
and introducing the auxiliary variables, $r$ and $s$, respectively by $+i(\hbar/2)\, \partial_q$ (cf. Eq~\eqref{Wigner222BB}) and $-i(\hbar/2)\, \partial_p$ (cf. Eq~\eqref{Wigner222CC}) one recovers an equivalent Wigner continuity equation cast in the form of Eq.~\eqref{alexquaz51}, with
\begin{equation}
J_q(q,\,p;\,t) = +\sum_{\eta=0}^{\infty} \left(\frac{i\,\hbar}{2}\right)^{2\eta}\frac{1}{(2\eta+1)!} \, \left[\partial_p^{2\eta+1} K(p)\right]\,\partial_q^{2\eta}W(q,\,p;\,t),
\label{alexquaz500BB}
\end{equation}
and
\begin{equation}
J_p(q,\,p;\,t) = -\sum_{\eta=0}^{\infty} \left(\frac{i\,\hbar}{2}\right)^{2\eta}\frac{1}{(2\eta+1)!} \, \left[\partial_q^{2\eta+1} V(q)\right]\,\partial_p^{2\eta}W(q,\,p;\,t),\label{alexquaz500CC}
\end{equation}
which, also from Eq.~\eqref{alexquaz51}, lead to an explicit form of the stationarity quantifier given by
\begin{equation} \label{helps}
\mbox{\boldmath $\nabla$}_{\xi} \cdot \mathbf{J} = -\partial_t W = \sum_{\eta=0}^{\infty}\frac{(-1)^{\eta}\hbar^{2\eta}}{2^{2\eta}(2\eta+1)!} \, \left\{
\left[\partial_p^{2\eta+1}K(p)\right]\,\partial_q^{2\eta+1}W
-
\left[\partial_q^{2\eta+1}V(q)\right]\,\partial_p^{2\eta+1}W
\right\}.\end{equation}
To fully capture the quantum distortions over the classical Hamiltonian regime described by currents in the form of \eqref{alexquaz500BB2}-\eqref{alexquaz500CC2}, the Liouvillianity quantifier (as in Eq.~\eqref{zeqnz59}) is expressed by
\begin{equation}
\mbox{\boldmath $\nabla$}_{\xi} \cdot \mathbf{w} = \sum_{\eta=1}^{\infty}\frac{(-1)^{\eta}\hbar^{2\eta}}{2^{2\eta}(2\eta+1)!}
\left\{
\left[\partial_p^{2\eta+1}K(p)\right]\,
\partial_q\left[\frac{1}{W}\partial_q^{2\eta}W\right]
-
\left[\partial_q^{2\eta+1}V(q)\right]\,
\partial_p\left[\frac{1}{W}\partial_p^{2\eta}W\right]
\right\} ~~~\end{equation}
which, together with Eq.~\eqref{helps}, encompasses all the contributions from quantum corrections of order $\mathcal{O}(\hbar^{2\eta})$.

\section{Dimensionless analysis for gaussian ensembles}

A more convenient way to depict the phase-space dynamics is through a dimensionless description of the Hamiltonian, $H^{W}(q,\,p)$ (cf. Eq.~\eqref{nlh}), i.e. through \cite{Novo21A,NossoPaper}
\begin{equation}
\label{dimHH}\mathcal{H}(x,\,k) = \mathcal{K}(k) + \mathcal{V}(x),
\end{equation}
written in terms of dimensionless variables, $x = \left(m\,\omega\,\hbar^{-1}\right)^{1/2} q$ and $k = \left(m\,\omega\,\hbar\right)^{-1/2}p$.

Assuming the above modifications, a gaussian distribution written as
\begin{equation}
\mathcal{G}_\alpha(x,\,k) = \hbar \,G_\alpha(q,\,p) = \frac{\alpha^2}{\pi}\, \exp\left[-\alpha^2\left(x^2+ k^2\right)\right],
\end{equation}
can be identified as the Wigner function so that the associated Wigner flow contributions assume the form
\begin{eqnarray}
\label{imWA2}\partial_x\mathcal{J}_x(x, \, k;\,\tau) &=& +\sum_{\eta=0}^{\infty} \left(\frac{i}{2}\right)^{2\eta}\frac{1}{(2\eta+1)!} \, \left[\partial_k^{2\eta+1}\mathcal{K}(k)\right]\,\partial_x^{2\eta+1}\mathcal{G}_{\alpha}(x, \, k),
\\
\label{imWB2}\partial_k\mathcal{J}_k(x, \, k;\,\tau) &=& -\sum_{\eta=0}^{\infty} \left(\frac{i}{2}\right)^{2\eta}\frac{1}{(2\eta+1)!} \, \left[\partial_x^{2\eta+1}\mathcal{V}(x)\right]\,\partial_k^{2\eta+1}\mathcal{G}_{\alpha}(x, \, k),
\end{eqnarray}
for the Hamiltonian, Eq.~(\ref{dimHH}).
From gaussian relations with Hermite polynomials of order $n$, $\mbox{\sc{H}}_{n}$, and assuming some properties for $\mathcal{V}$ and $\mathcal{K}$ derivatives\footnote{That is, when $\mathcal{V}$ and $\mathcal{K}$ derivatives can be eventually cast in the form of
\begin{eqnarray}
\label{t111}
\partial_x^{2\eta+1}\mathcal{V}(x) &=& \lambda^{2\eta+1}_{(x)} \, \upsilon(x),\nonumber\\
\label{t222}
\partial_k^{2\eta+1}\mathcal{K}(k) &=& \mu^{2\eta+1}_{(k)} \, \kappa(k),\nonumber
\end{eqnarray}
with $\lambda$, $\upsilon$, $\mu$, and $\kappa$ identified as arbitrary auxiliary functions.} \cite{Novo21A}, after some straightforward mathematical manipulations \cite{Gradshteyn,Novo21A,NossoPaper},
one obtains
\begin{eqnarray}
\label{imWA3}\partial_x\mathcal{J}_x(x, \, k;\,\tau) &=& (+2i) \kappa(k)\,\mathcal{G}_{\alpha}(x, \, k)\,\sum_{\eta=0}^{\infty} \left(\frac{i\,\alpha\,\mu_{(k)}}{2}\right)^{2\eta+1}\frac{1}{(2\eta+1)!} \, \mbox{\sc{H}}_{2\eta+1} (\alpha x),\\
\label{imWB3}\partial_k\mathcal{J}_k(x, \, k;\,\tau) &=& (-2i) \upsilon(x)\,\mathcal{G}_{\alpha}(x, \, k)\sum_{\eta=0}^{\infty} \left(\frac{i\,\alpha\, \lambda_{(x)}}{2}\right)^{2\eta+1}\frac{1}{(2\eta+1)!} \, \mbox{\sc{H}}_{2\eta+1} (\alpha k),\end{eqnarray}
which, for a convergent series, result in the stationarity quantifier, $\mbox{\boldmath $\nabla$}_{\xi}\cdot \mbox{\boldmath $\mathcal{J}$}$, of the associated Wigner flow.

\section{Stationary gaussian ensembles -- a quantum camouflage}

Squeezed gaussian ensembles are introduced as\footnote{The correspondence with the physical coordinates, $q$ and $p$, is obtained from
\begin{equation}
G_\zeta(q,\,p) = \frac{1}{\pi\hbar}\, \exp\left[-\frac{1}{\hbar}\left(e^{+2\zeta}\frac{q^2}{\mathcal{A}^2}+ e^{-2\zeta}\mathcal{A}^2\,p^2\right)\right],
\end{equation}
which yields back $\mathcal{G}_\zeta(x,\,k) = \hbar \,G_\zeta(q,\,p) $ with $\mathcal{A} = (m\,\omega)^{-1}$, for mass scale, $m$, and angular frequency, $\omega$.}
\begin{equation}
\mathcal{G}_\zeta(x,\,k) = \frac{1}{\pi}\, \exp\left[-\left(e^{+2\zeta}x^2+ e^{-2\zeta}k^2\right)\right],
\end{equation}
which replaces the Wigner function, in Eqs.~\eqref{imWA3} and \eqref{imWB3}, with $\alpha \to \zeta$, so to lead to the following associated gaussian flow contributions,
\begin{eqnarray}
\label{imWA22}\partial_x\mathcal{J}^{\zeta}_x(x, \, k;\,\tau) &=& +\sum_{\eta=0}^{\infty} \left(\frac{i}{2}\right)^{2\eta}\frac{1}{(2\eta+1)!} \, \left[\partial_k^{2\eta+1}\mathcal{K}(k)\right]\,\partial_x^{2\eta+1}\mathcal{G}_{\zeta}(x, \, k),
\\
\label{imWB22}\partial_k\mathcal{J}^{\zeta}_k(x, \, k;\,\tau) &=& -\sum_{\eta=0}^{\infty} \left(\frac{i}{2}\right)^{2\eta}\frac{1}{(2\eta+1)!} \, \left[\partial_x^{2\eta+1}\mathcal{V}(x)\right]\,\partial_k^{2\eta+1}\mathcal{G}_{\zeta}(x, \, k).
\end{eqnarray}
For a Hamiltonian given by $\tilde{\mathcal{H}}(x,\,k) = \tilde{\mathcal{K}}(k) + \tilde{\mathcal{V}}(x)$, with
\begin{eqnarray}\label{seila}
\tilde{\mathcal{V}}(x) &=& \lambda_1\cosh(\nu_1\,x) + \lambda_2 \,\cos(\nu_2\,x),\\
\tilde{\mathcal{K}}(k) &=& \gamma_1\cosh(\mu_1\,k) + \gamma_2 \,\cos(\mu_2\,k),
\end{eqnarray}
where $\lambda_i$ and $\gamma_i$ ($i=1,\,2$) are arbitrary constants, and for gaussian derivatives rewritten as \cite{Novo21A},
\begin{equation}\label{ssae}
\partial_x^{2\eta+1}\mathcal{G}_{\zeta}(x, \, k) = (-1)^{2\eta+1}e^{+(2\eta+1)\zeta}\,\mbox{\sc{H}}_{2\eta+1} (e^{+ \zeta} x)\, \mathcal{G}_{\zeta}(x, \, k),
\end{equation}
\begin{equation}\label{ssaf}
\partial_k^{2\eta+1}\mathcal{G}_{\zeta}(x, \, k) = (-1)^{2\eta+1}e^{-(2\eta+1)\zeta}\,\mbox{\sc{H}}_{2\eta+1} (e^{- \zeta} k)\, \mathcal{G}_{\zeta}(x, \, k),
\end{equation}
one can verify that, besides the classical stationarity (as used to be associated to thermodynamic ensembles \cite{Novo21A}), the quantum-driven gaussian ensembles, $\mathcal{G}_{\zeta}(x,\,k;\,\tau)$, restores the Wigner flow stationary regime, in a kind of a quantum {\em camouflage} of the classical pattern.

Through some straightforward mathematical manipulations, and introducing the constraint $\mu_{1(2)} = e^{-2\zeta}\nu_{2(1)}$, one has $\mbox{\boldmath $\nabla$}_{\xi} \cdot \mbox{\boldmath $\mathcal{J}^{\zeta}$} = \partial_x\mathcal{J}^{\zeta}_x+\partial_k\mathcal{J}^{\zeta}_k$ re-written as
\begin{eqnarray}
\label{imWA22mmBB}\mbox{\boldmath $\nabla$}_{\xi} \cdot \mbox{\boldmath $\mathcal{J}^{\zeta}$} &=&
+2\left[
\sin(\mu_2\,k)\,\sinh(\nu_1\,x)\,\left(\gamma_2\, e^{-\frac{\nu_1\mu_2}{4}}+\lambda_1\, e^{+\frac{\nu_1\mu_2}{4}}\right)\right.\nonumber\\&&
\qquad\qquad\qquad\left. - \sinh(\mu_1\,k)\,\sin(\nu_2\,x)\,\left(\gamma_1 \,e^{+\frac{\nu_2\mu_1}{4}}+\lambda_2 \,e^{-\frac{\nu_2\mu_1}{4}}\right)
\right]
\mathcal{G}_{\zeta}(x, \, k),\,\,\,\,\quad
\end{eqnarray}
from which the stationary behavior, $\mbox{\boldmath $\nabla$}_{\xi} \cdot \mbox{\boldmath $\mathcal{J}^{\zeta}$}=0$, is recovered for
$$\lambda_{2(1)} = -\gamma_{1(2)}\,\exp[+(-)\nu_{2(1)}\mu_{1(2)}/2],$$
which sets the Hamiltonian, $\tilde{\mathcal{H}}(x,\,k)$, with $\tilde{\mathcal{K}}(k)$ and $\tilde{\mathcal{V}}(x)$ re-written in the simplified form
\begin{eqnarray}\label{seilaB}
\tilde{\mathcal{V}}(x) &=& -\exp[-e^{-2\zeta}/2]\cosh(x) - \,\gamma\cos(e^{+2\zeta}\,x),\\
\label{seila2B}\tilde{\mathcal{K}}(k) &=& +\gamma\exp[-e^{+2\zeta}/2]\cosh(k) + \cos(e^{-2\zeta}\,k),
\end{eqnarray}
and the quantum pattern is {\em camouflaged} by the stationarity of the gaussian ensemble.

\section{Conclusions}

The phase-space WW framework for investigating classical to quantum transition of systems driven by $1$-dim non-linear equations of motion encompassed by a Hamiltonian dynamics described by $\mathcal{H}(x,\,k)$, under the condition $\partial^2 \mathcal{H} / \partial x \, \partial k = 0$, was discussed and specified for to the study of gaussian ensembles.
As evinced by our calculations, even through the Schr\"odinger-like equation, the gaussian ensemble, $\mathcal{G}_{\zeta}(x, \, k)$, is just the zero-mode of $\tilde{\mathcal{H}}(x, \, k)$, i.e. $\tilde{\mathcal{H}}\,\mathcal{G}_{\zeta} = 0$, which can be the first step for obtaining the complete spectrum of exotic Hamiltonians like $\tilde{\mathcal{H}}(x, \, k)$.

The answer for this question is not only concerned with the form of $\tilde{\mathcal{V}}(x)$ and $\tilde{\mathcal{K}}(k)$, Eqs.~\eqref{seilaB}-\eqref{seila2B}, but also with the role of quantum gaussian ensembles in yielding back well-behaved quantum current descriptions of non-standard Hamiltonian systems by rendering its stationary behavior.
Our investigation was concerned with a specific class of Hamiltonians for which gaussian ensembles were identified as the quantum ground state, camouflaging quantum effects so as to exhibit a stationary flow behavior.

Of course, our analysis do not exhaust all the possible algorithms and can be regarded as a preliminary step towards a more general and realistic model to describe quantum effects in non-linear systems.

\vspace{.5 cm}
{\em Acknowledgments -- The work of AEB is supported by the Brazilian Agencies FAPESP (Grant No. 2023/00392-8, S\~ao Paulo Research Foundation (FAPESP)) and CNPq (Grant No. 301485/2022-4).}


\begin{thebibliography}{99}
\bibitem{Wigner}
E. Wigner, Phys. Rev. {\bf 40} 749 (1932).
\bibitem{Ballentine}
L. E. Ballentine, {\em Quantum Mechanics: a Modern Development}, pp. 633 (World Scientific, 1998).
\bibitem{Case}
W. B. Case, Am. J. Phys. {\bf 76}, 937 (2008).
\bibitem{Meu2018}
A. E. Bernardini, Phys. Rev. A {\bf 98}, 052128 (2018).
\bibitem{Neumann}
J. von Neumann, {\em Mathematical Foundations of Quantum Mechanics}, Translated by R. T. Beyer, (Princeton University Press, 1955).
\bibitem{Zurek01}
W. H. Zurek, Phys. Rev. D {\bf 24}, 1516 (1981); Phys. Rev. D {\bf 26}, 1862 (1982).
\bibitem{Zurek02}
W. H. Zurek, Phys. Today {\bf 44}, 36 (1991).
\bibitem{Steuernagel3}
O. Steuernagel, D. Kakofengitis and G. Ritter, Phys. Rev. Lett. {\bf 110}, 030401 (2013).
\bibitem{Novo21A}
A. E. Bernardini and O. Bertolami, Phys. Rev. {\bf A105}, 032207 (2022).
\bibitem{Novo21B}
A. E. Bernardini and O. Bertolami, Phys. Rev. {\bf E106}, 024202 (2022). 
\bibitem{Novo21B2}
A. E. Bernardini and O. Bertolami, Phys. Rev. {\bf E107}, 044201 (2023).
 \bibitem{NossoPaper}
A. E. Bernardini and O. Bertolami, EPL {\bf 120}, 20002 (2017); A. E. Bernardini and O. Bertolami, Journal of Physics: Conf. Series {\bf 1275}, 012032 (2019).
\bibitem{Gradshteyn}
I. S. Gradshteyn and I. Ryzhik, {\it Tables of Integrals, Series and Products} (Academic Press, New York, 1994).
\end{thebibliography}
\end{document}